\documentclass[onecolumn,11pt]{revtex4}
\usepackage{graphicx,epsf}
\usepackage{color}
\usepackage{dcolumn}
\usepackage{mathrsfs}

\usepackage{subfig}
			
\begin{document}

\title{On fast radial propagation of parametrically excited geodesic acoustic mode}

\author{Z. Qiu$^{1}$, L. Chen$^{1, 2}$ and F. Zonca$^{3, 1}$}

\affiliation{$^1$Institute for    Fusion Theory and Simulation and Department of Physics, Zhejiang University, Hangzhou, P.R.C\\
$^2$Department of   Physics and Astronomy,  University of California, Irvine CA 92697-4575, U.S.A.\\
$^3$ ENEA C. R. Frascati, C. P.
65-00044 Frascati, Italy}

\begin{abstract}
The spatial and temporal evolution of parametrically excited geodesic acoustic mode (GAM) initial pulse is investigated both analytically and numerically.
Our results show that  the nonlinearly excited GAM  propagates at a   group velocity which is, typically, much larger than that due to finite ion Larmor radius as predicted by the linear theory. The nonlinear dispersion relation of GAM driven by a  finite amplitude drift wave pump is also derived, showing a nonlinear frequency increment of GAM.
Further implications of these findings for  interpreting  experimental observations are also discussed.
\end{abstract}

\maketitle

\section{Introduction}

Geodesic Acoustic Modes (GAM) \cite{NWinsorPoF1968,FZoncaEPL2008} are finite-frequency components of zonal structures (ZS) \cite{AHasegawaPoF1979,MRosenbluthPRL1998}
unique in toroidal plasmas, which could be   spontaneously excited by  microscopic drift wave (DW) type
turbulence \cite{WHortonRMP1999},   including drift Alfv\'en waves (DAW); and in turn, are  capable of scattering DW/DAW into stable short radial-wavelength domain \cite{FZoncaEPL2008,LChenPRL2012,ZQiuEPL2013,ZQiuPoP2014,ZLinScience1998}. Therefore, GAM may regulate the
turbulence intensity and the associated wave-induced transports \cite{PLiewerNF1985,ZFengPoP2013}.   The excitation of GAM by DWs   can be described by parametric decay instability \cite{RSagdeevbook1969,PKawPoF1969}, where the DWpump    $(\omega_0, \mathbf{k}_0)$ decays into a DW lower sideband $(\omega_S, \mathbf{k}_S)$ and a GAM $(\omega_G, \mathbf{k}_G)$, and   selection rules of frequency and wavenumber matching conditions are satisfied.

It is known that GAM has a finite linear group velocity due to finite ion Larmor radius (FILR) effects, and this linear group velocity is typically radially outward, consistent with GAM continuum  due to radial temperature profiles. For GAM with $k_r\rho_i\ll1$, its linear group velocity is   $V_G=C_G\omega_G k_r\rho^2_i$, with the linear dispersion relation given as $\omega^2=\omega^2_G(1+C_Gk^2_r\rho^2_i)$ \cite{FZoncaEPL2008}. Here, $\omega_G=\sqrt{T_e/T_i+7/4} v_{ti}/R_0$ is the local GAM frequency in the fluid limit, $v_{ti}\equiv\sqrt{2T_i/m_i}$ the thermal ion velocity, $R_0$ is the major radius of the torus, $k_r$ is the radial wavevector, $\rho_i=m_icv_{ti}/(eB_0)$ is the ion Larmor radius, and $C_G$ is an order unity coefficient.
This linear group velocity of GAM  has been discussed in several works
\cite{FZoncaEPL2008,ZQiuPST2011,RHagerPoP2012}, and is shown to have important consequences on the nonlinear excitation of GAM by DWs  and change the absolute/convective nature of the parametric instability \cite{LChenVarenna2010,ZQiuPoP2014}.
Radial propagation of GAM has been observed in several experiments \cite{TIdoPPCF2006,GConwayPPCF2008,DKongNF2013},   and qualitative agreement between the experimental results and linear theory has been obtained  \cite{YHamadaNF2011,DKongNF2013}.
However, in-depth analysis of the experimentally obtained dispersion relation leads to the conclusion that, even though a quadratic dependence of GAM frequency on its radial wavevector, qualitatively  consistent with linear theory of KGAM \cite{DKongNF2013}, is indeed obtained,  the   coefficient for   FILR effects is much larger than that  predicted by linear theory \cite{ZQiuPPCF2009,ZQiuPST2011}. This discrepancy has also been found in   numerical simulations \cite{RHagerPoP2012}, where the measured radial propagation velocity of the DW driven   GAM is   used to determine the coefficients of FILR effect, and is found to be  much larger than unity.
Up to now, there is no  first-principle-theory-based explanation of this  ``enhanced FILR effect",   although a general theoretical framework exists to formulate it with all necessary physics ingredients \cite{ZQiuPoP2014}. In fact,  we will   show in this work, that this discrepancy could be due to nonlinear effects.

We note  that, GAM is an $n=0/m\simeq0$ mode, with $m=\pm 1$ sidebands and possibly higher order ones (depending on the perpendicular wavelength \cite{FZoncaEPL2008}),  such that it is generally not driven unstable by   expansion free energy of the plasma. Here, $m$ and $n$  are, respectively, the poloidal and toroidal mode numbers in the Fourier mode structure representation adopting straight field line toroidal flux coordinates \cite{ABoozerPoF1981}. Thus, GAM, in general, could be   observed   when it is nonlinearly driven by ambient turbulence, and in this case, the spatial-temporal evolution of GAM  is significantly affected by the DW nonlinear drive. It has been pointed out in Ref. \citenum{ZQiuPoP2014} that  the nonlinearly driven GAM propagates at a much larger nonlinearly-coupled group velocity in the presence of DW.  As a result, the propagation of GAM  and experimental observations should also be interpreted taking nonlinear effects into account.  In this work, we shall further study the spatial-temporal evolution of the nonlinearly coupled DW-GAM system more in   details in order to analyze its implications to experimental observations.  The rest of the paper is organized as follows: the theoretical model used in this work is presented in Sec. \ref{sec:model}, which is then solved in Sec. \ref{sec:evolution} for the spationtemporal evolution of the coupled GAM-DW system. The possible applications of our theory to interpretation of experimental observations and numerical results   are also  presented. Finally, a brief summary is given in Sec. \ref{sec:summary}.

\section{Theoretical Model}\label{sec:model}

The  equations describing the nonlinear interactions between GAM and DW   are derived using gyrokinetic theory \cite{FZoncaEPL2008,ZQiuPoP2014}. Assuming that the DW  is constituted by  a constant-amplitude pump wave and a lower sideband with a much smaller amplitude due to GAM modulation,   the normalized coupled nonlinear equations describing GAM   excitation by DW are then given as equations (9) and (10) of Ref. \citenum{ZQiuPoP2014}:
\begin{eqnarray}
\left(\partial_t+\gamma_S+i\omega_P-i\omega_*-iC_d\omega_*\rho^2_i\partial^2_r\right)A_S&=&\Gamma^*_0\mathscr{E},\label{DWSBequation}\\
\left(\partial_t(\partial_t+2\gamma_G)+\omega^2_G -C_G\omega^2_G \rho^2_i \partial^2_r\right)\mathscr{E}&=&i\omega_G\Gamma_0\partial^2_rA_S.\label{GAMequation}
\end{eqnarray}
Here, $\mathscr{E}\equiv \partial_r A_G/\alpha$ is related with the GAM electric field with $\alpha\equiv i(\alpha_i\omega_PT_e/T_i)^{1/2}$ and $\alpha_i\equiv1+\delta P_{\perp}/(en_0\delta\phi_P)$   an order unity coefficient \cite{LChenPoP2000}, $\delta P_{\perp}$ is the perturbed perpendicular pressure due to the DW pump scalar potential $\delta\phi_P$ in the $k_{\perp}\rho_i\ll1$ limit \cite{LChenPoP2000}. Meanwhile,  $A_G$, $A_P$ and $A_S$ are, respectively,  the radial envelopes of GAM, DW pump and lower sideband,  $\Gamma_0\equiv (\alpha_i T_i/\omega_PT_e)^{1/2}ck_{\theta,P}  A_P/B$ is the normalized pump wave amplitude. Furthermore,   $\gamma_S$ and $\gamma_G$ are the Landau damping rates of DW sideband and GAM, $\omega_P$ is the   pump DW frequency and  $\omega_*$ is the diamagnetic drift frequency. The kinetic term in equation (\ref{DWSBequation}), i.e., the term proportional to $C_d$, comes from finite radial envelope variation due to the coupling between neighboring poloidal harmonics. The expression for $C_d$  can be derived from equation (19) of Ref. \citenum{FRomanelliPoFB1993}, and one has  $C_d\sim O(\epsilon/(n^2q'^2\rho^2_i))$ with $q$ being the safety factor, $q'=dq/dr$ its radial derivative  and $\epsilon=r/R_0$. On the other hand, the kinetic term in equation (\ref{GAMequation}); i.e.,  the term proportional to $C_G$, comes from   FILR of GAM. Thus, $C_G\sim O(1)$ and its detailed expression can be obtained from equation (9) of Ref. \cite{ZQiuPPCF2009}. Other notations are standard. We note that, the governing equations (\ref{DWSBequation}) and (\ref{GAMequation}) are derived from quasi-neutrality condition assuming both GAM and DW are predominantly electrostatic perturbations. Electrons respond  adiabaticly to $k_{\parallel}\neq0$ perturbations, i.e.,  DW and $m\neq0$ poloidal sidebands of GAM; while ion responses are solved assuming $q\gg1$, $k_{\perp}\rho_i\ll1$ and $|\omega_0|\sim|\omega_*|$ for DW.  
We note that, even though turbulence usually refers to a broad spectrum of nonlinearly interacting DWs, in the present analysis we  have considered the nonlinear interactions of GAM with a single-$n$ DW in order to elucidate the  nonlinear effects on the radial propagation of GAM due to interaction with finite-amplitude DWs.  Since for each DW with toroidal mode number $n$, the interactions with the corresponding GAM  is coherent, we may expect  that  in the presence of DW turbulence consisting of multiple-$n$ modes, the net nonlinear effects would be an appropriate sum/integral of the nonlinear effects of individual-$n$ mode sconsidered here.
System nonuniformities   in equations (\ref{DWSBequation}) and (\ref{GAMequation}), which may affect qualitatively the convective/absolute nature of the parametric process  as shown in Ref. \citenum{ZQiuPoP2014}, are also  ignored here in order to  focus on the radial propagation of the parametrically  excited GAM pulse.

\section{Spatialtemporal evolution of the coupled DW-GAM system}\label{sec:evolution}

Equations (\ref{DWSBequation}) and (\ref{GAMequation}) can be solved using two-spatial two-temporal scales expansion of $\mathscr{E}$ and $A_S$, i.e., $A_S=\hat{A}_S(\tau,\zeta)\exp(ik_0r-i\omega_0t)$ and $\mathscr{E} =\hat{\mathscr{E}}(\tau,\zeta)\exp(ik_0r-i\omega_0t)$ such that $\partial_t=-i\omega_0+\partial_{\tau}$ and $\partial_r=i k_0+\partial_{\zeta}$, with $\tau$ and $\zeta$ denoting the slow temporal and spatial variations. In order to delineate the physics of GAM propagation, we can assume that the system is well above the excitation threshold \cite{FZoncaEPL2008}. Thus, we can ignore $\gamma_S$ and $\gamma_G$ in equations (\ref{DWSBequation}) and (\ref{GAMequation}); and the coupled nonlinear equations reduce to
\begin{eqnarray}
\left(\partial_{\tau} +V_S\partial_{\zeta}\right)\hat{A}_S&=&\Gamma^*_0\hat{\mathscr{E}},\label{DWSBreduced}\\
\left(\partial_{\tau} +V_G\partial_{\zeta}\right)\hat{\mathscr{E}}&=&\frac{1}{2}\Gamma_0(k^2_0-2ik_0\partial_{\zeta})\hat{A}_S. \label{GAMreduced}
\end{eqnarray}
Here, $V_S=2C_d\omega_*\rho^2_ik_0$ and $V_G=C_G\omega_G\rho^2_ik_0$ are, respectively, the linear group velocities of DW sideband and GAM. We note that  $V_S$ and $V_G$ have the same sign for typical tokamak parameters \cite{FZoncaEPL2008,ZQiuPoP2014}, such that the excitation of GAM by DWs is a convective amplification process, ignoring system nonuniformities \cite{MRosenbluthPRL1972,LChenVarenna2010}. Furthermore, in deriving equations (\ref{DWSBreduced}) and (\ref{GAMreduced}), the following frequency and wavenumber matching conditions for resonant decay are applied
\begin{eqnarray}
-\omega_0+\omega_P-\omega_*+C_d\omega_*k^2_0\rho^2_i&=&0,\nonumber\\
-\omega_0^2+\omega^2_G+C_G\omega^2_Gk^2_0\rho^2_i&=&0,\nonumber
\end{eqnarray}
from where  $(\omega_0,k_0)$  can be solved for.

Moving into the wave frame by taking $\xi=\zeta-V_c\tau$, with $V_c=(V_S+V_G)/2$, the coupled nonlinear equations, (\ref{DWSBreduced}) and (\ref{GAMreduced}),
can  be combined to yield the following equation describing the nonlinear spatialtemporal evolution  of the parametrically excited GAM
\begin{eqnarray}
\left(\partial^2_{\tau}-V^2_0\partial^2_{\xi}\right)\hat{\mathscr{E}}=\frac{1}{2}k^2_0\Gamma^2_0\hat{\mathscr{E}}-ik_0
\Gamma^2_0\partial_{\xi}\hat{\mathscr{E}}.\label{GAMequation1}
\end{eqnarray}
Here, $V_0=(V_S-V_G)/2$. Letting $\hat{\mathscr{E}}=\exp(i\beta\xi)A(\xi,\tau)$, with $\beta=k_0\Gamma^2_0/(2V^2_0)$, equation (\ref{GAMequation1}) reduces to
\begin{eqnarray}
\left(\partial^2_{\tau}-V^2_0\partial^2_{\xi}\right)A=\left(\frac{1}{2}k^2_0\Gamma^2_0+\beta k_0\Gamma^2_0-\beta^2V^2_0\right) A
\equiv \hat{\eta}^2A.\label{GAMequation2}
\end{eqnarray}
Physically, $\beta$ can be interpreted as nonlinear modification to the GAM wave vector, which also affects the GAM frequency as shown below. Equation (\ref{GAMequation2}) can be solved, and yields the following unstable solution
\begin{eqnarray}
A=\frac{\hat{A}_0}{\sqrt{\pi}\Delta k_0}\int^{\infty}_{-\infty}dk_I\exp\left(-\frac{k^2_I}{\Delta k^2_0}\right)\exp\left[ik_I\xi+\sqrt{\hat{\eta}^2-k^2_IV^2_0}\tau\right].\label{generalsolution}
\end{eqnarray}

This solution corresponds to the initial condition
\begin{eqnarray}
A=\hat{A}_0\exp\left(-\frac{\Delta k^2_0\xi^2}{4}\right)\nonumber
\end{eqnarray}
at $\tau=0$. We note that this is the typical wave packet initial structure   for parametrically excited GAM, with a spectrum width $\Delta k_0$. Assuming $|V_c\partial_{\xi}|\ll|\partial_{\tau}|$, i.e., convective damping due to FILR effects are higher order corrections to the temporal growth \cite{FZoncaEPL2008,NChakrabartiPoP2008}, the general solution, equation (\ref{generalsolution}), can then be reduced to the following time asymptotic solution:
\begin{eqnarray}
A=\frac{\hat{A}_0}{\Delta k_0\lambda_{\tau}}\exp\left(\hat{\eta}\tau-\frac{\xi^2}{4\lambda^2_{\tau}}\right).
\end{eqnarray}
Here, $\lambda^2_{\tau}=(1/\Delta k^2_0+V^2_0\tau/(2\hat{\eta}))$, and it corresponds to GAM initial pulse broadening in time. The time asymptotic solution of GAM electric field is then
\begin{eqnarray}
\mathscr{E}=\frac{\hat{A}_0}{\Delta k_0\lambda_{\tau}}\exp\left(\hat{\eta}\tau+i\beta(\zeta-V_c\tau)-\frac{1}{4\lambda^2_{\tau}}\left(\zeta-V_c\tau\right)^2\right).\label{asymptoticsolution}
\end{eqnarray}

One then   readily has from equation (\ref{asymptoticsolution}) that the nonlinearly excited GAM is characterized by a nonlinear radial wavevector
\begin{eqnarray}
k_{NL}=k_0-i\partial_{\zeta}\ln\mathscr{E}=k_0\left(1+\Gamma^2_0/(2V^2_0)\right),\label{nonlinearwavevector}
\end{eqnarray}
i.e., the wavevector increases with pump DW amplitude, and is  larger than that predicted from frequency/wavenumber matching conditions.

The real frequency of the excited GAM   can also be obtained from equation (\ref{asymptoticsolution})
\begin{eqnarray}
\omega_{NL}=\omega_0+i\partial_\tau\ln\mathscr{E}=\omega_0+\frac{k_0\Gamma^2_0V_c}{2V^2_0}.\label{nonlinearfrequency}
\end{eqnarray}
$\omega_0(k_0)$ can be solved from the matching conditions, which can then be substituted into equation (\ref{nonlinearfrequency}), and yield:
\begin{eqnarray}
\omega_{NL}&=&\omega_G+\frac{k_0\Gamma^2_0V_c}{2V^2_0}+\frac{1}{2}C_G\omega_Gk^2_0\rho^2_i\nonumber\\
&=&\omega_G+\frac{k_0\Gamma^2_0V_c}{2V^2_0}+\frac{C_G\omega_G\rho^2_ik^2_{NL}}{2(1+\Gamma^2_0/(2V^2_0))^2}.\label{nonlinearDR}
\end{eqnarray}
This is the nonlinear dispersion relation of the parametrically excited GAM.
We note that, both $V_0$ and $V_c$ are proportional to $k_0$, and thus, the nonlinear frequency shift due to the modulation of DW, $k_0\Gamma^2_0V_c/(2V^2_0)$, is independent of $k_0$. Thus, finite amplitude DW will increase the frequency of the nonlinearly driven GAM. The frequency increment, can be expressed as $(e\delta\phi/T)^2(L_n/\rho_i)^2$ from our theory, which indicates an order of unity frequency increment for    typical tokamak parameters. This may explain the existence of the higher frequency branch of    the  ``dual-GAM" observed in HT-7 tokamak \cite{DKongNF2013}, which oscillates at a frequency much higher than other branch with the usual GAM frequency (The frequencies of the two co-existing ``dual - GAMs" are respectively 12 and 21 kHz in shot 113901 \cite{DKongNF2013}). Another finding of the HT-7 experiment is the coefficient of FILR effect is $O(10^2)$ larger than that predicted by linear theory \cite{DKongprivate}.
On the other hand, equation (\ref{nonlinearDR}) shows that the coefficient for kinetic dispersiveness is, in fact, decreased by a factor   $(1+\Gamma^2_0/(2V^2_0))^2$. The reason why experimental analysis found an ``increased" coefficient is that, in the analysis of experimental data, one employed the linear dispersion relation of GAM and used the expression $(\omega_{obs}-\omega_{loc})/(\omega_{loc}k^2_{obs}\rho^2_i)$ to determine the coefficient  $C_G$ \cite{DKongprivate}. Here, the subscript $``obs"$ denotes experimental observation, and $``loc"$ denotes local continuum frequency of GAM. As we have shown in equation (\ref{nonlinearDR}),  ``$\omega_{obs}-\omega_{loc}$" contains  the kinetic dispersiveness as well as the order one nonlinear frequency increment $k_0\Gamma^2_0V_c/(2V^2_0)$; which, thus, can  lead to an over-estimation of the coefficient $C^{NL}_G$ \cite{DKongprivate}. The effective coefficient obtained in this way is, $C^*_G\sim(e\delta\phi/T)^2(L_n/\rho_i)^2/(k^2_G\rho^2_i)\sim O(10^2)$ for typical tokamak parameters; which is  significantly larger than that predicted by linear theory. Our nonlinear theory, thus, provides a possible explanation of experimental observations. It can also be used to explain the $O(10^2)$ increase of the FILR coefficient from numerical simulations \cite{RHagerPoP2012}.

The coupled GAM and DW sideband wavepacket,  propagates at a nonlinear group velocity $V_c=(V_S+V_G)/2$, which is much larger than the linear group velocity of GAM due to $|V_S|\gg |V_G|$ ($|\omega_P|\simeq |\omega_*|\gg|\omega_G|$ for resonant decay). Thus, to interpret the propagation of GAM nonlinearly excited by DW turbulences including DAW, linear theory of KGAM \cite{FZoncaEPL2008,ZQiuPPCF2009} is not adequate, and one must instead, apply  nonlinear theory. We note also that, while both the real frequency and wavevector of the excited GAM depend on the amplitude of the pump DW, the nonlinear group velocity is determined by $k_0$ from matching conditions, and is independent of the pump amplitude. Thus, for the comparison of experimentally observations with analytical theory, the nonlinear group velocity may be a better candidate.

\begin{figure}
\setlength{\unitlength}{1cm}
\begin{center}
\begin{minipage}[t]{9.5cm}
\includegraphics[width=9cm]{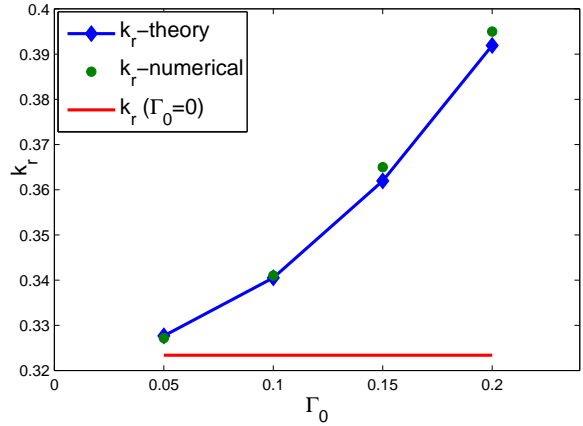}
\caption{ Nonlinear wavenumber $k_r$ v.s. pump amplitude $\Gamma_0$ }\label{fig:nonlinearkr}
\end{minipage}
\end{center}
\end{figure}

\begin{figure}
\setlength{\unitlength}{1cm}
\begin{center}
\begin{minipage}[t]{9.5cm}
\includegraphics[width=9cm]{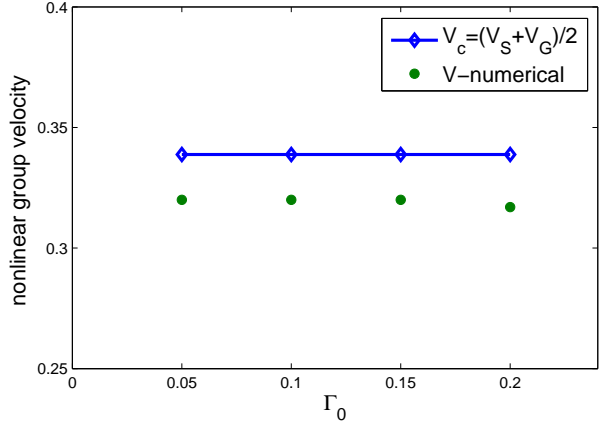}
\caption{ Nonlinear group velocity $v_g$ v.s. pump amplitude $\Gamma_0$ }\label{fig:nonlinearvg}
\end{minipage}
\end{center}
\end{figure}

The coupled nonlinear GAM and DW sideband equations, equations (\ref{DWSBequation}) and (\ref{GAMequation}), are solved numerically. Here, we fix  $C_d=C_G=1$, $\omega_G=0.1$, $\omega_P=\omega_*=1$, and study the coupled nonlinear equations by varying $\Gamma_0$.  The dependence of the nonlinear wavenumber $k_r$ on pump amplitude is given in Fig. \ref{fig:nonlinearkr}, where the dots are the wavenumbers from numerical solution, the diamonds  are the wavenumbers obtained from  equation (\ref{nonlinearwavevector}); and the solid curve is obtained from matching condition. For the parameters we have here, the wavevector   solved from matching conditions is $k_0=0.32$.  We may see from Fig. \ref{fig:nonlinearkr} that our nonlinear theory fits well with the numerical results; and it reduces to $k_0$ as $\Gamma_0$ approaches 0.  The comparison of the numerically measured nonlinear group velocity with our theory, is presented in Fig. \ref{fig:nonlinearvg}, where the dots are numerical results and the diamonds are obtained from $V_c=(V_S+V_G)/2$, and  $V_S$ and $V_G$ are defined with $k_0$. We note that, for the parameters we used in numerical solution, $V_S=0.64$, $V_G=0.032$ and $V_c=(V_S+V_G)/2=0.34\gg V_G$. Very good agreement between numerical results and analytical theory ($<3\%$ discrepency) are obtained here, suggesting that experimentally observed radial propagation of GAM must be understood using nonlinear theory.

\begin{figure}
\setlength{\unitlength}{1cm}
\begin{center}
\begin{minipage}[t]{9.5cm}
\includegraphics[width=9cm]{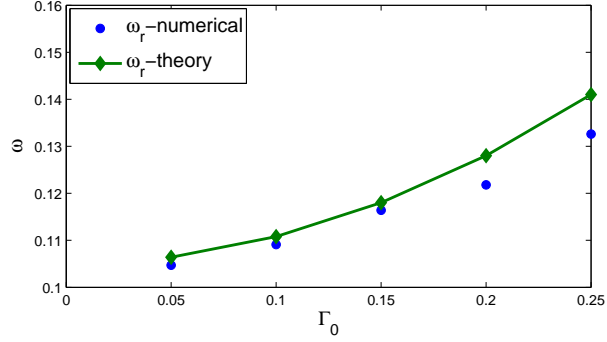}
\caption{ Nonlinear GAM frequency $\omega^{NL}$ v.s. pump amplitude $\Gamma_0$ }\label{fig:nonlinearfrequency}
\end{minipage}
\end{center}
\end{figure}

The nonlinear frequency of GAM is given in Fig. \ref{fig:nonlinearfrequency}, where the dots are numerical results and the diamonds represents $\omega_{NL}$ from equation (\ref{nonlinearfrequency}). Note that, for the parameters we use here, $\omega_0=0.105$, and the nonlinear frequency from numerical solution increases with pump DW amplitude as predicted by our theory.

\section{Conclusions and Discussions}\label{sec:summary}

In conclusion,   equations describing the spatialtemporal evolution of parametrically excited GAM initial pulse are studied both analytically and numerically. It is found that the parametrically excited GAM propagates at a nonlinear group velocity, which is the mean of the linear group velocities of GAM and DW, and is much larger than that predicted by linear theory of kinetic GAM. The wavevector of the excited GAM  has a quadratic dependence on the amplitude of the constant-amplitude pump DW. On the other hand, the nonlinear group velocity is independent of the pump DW amplitude; suggesting it as a good candidate for the comparison  between experiments and analytical theory. Our nonlinear theory, further shows that  there is a  nonlinear upshift in the GAM frequency.
Implications of the present theoretical findings to the HT-7 experimental observations are also important. Our results demonstrate that one must include nonlinear effects in order to properly analyze numerical simulations and/or experimental observations of GAM.

We note  that, while the ambient   turbulence in experiments consists of a whole spectrum of nonlinearly interacting DWs,  we have, in the present analysis,  considered the modification of the GAM dispersion relation due to a single-$n$  DW with finite amplitude. This  can be justified, since DW interactions with ZS, e.g. GAMs, have two components: a coherent part due to the interaction with the self-generated ZS, and a random contribution  due to interaction with ZS produced by other incoherent components of the fluctuation spectrum \cite{FZoncaNJP2015}. In both cases, the coupling coefficient is proportional to DW intensity and, therefore, we focus  here on the coherent GAM-DW interaction \cite{FZoncaEPL2008}. Effects  of system nonuniformities, which are shown to play important roles on the nonlinear interactions between GAM and DWs  are also  ignored here. This will limit the applications of our nonlinear theory.  To properly interprete global numerical simulations and/or experimental results, more in-depth investigations taking into account system nonunifomities will be needed.

\section*{Acknowledgments}
This work is supported by     US DoE GRANT,  the ITER-CN under Grants Nos. 2011GB105001,
2013GB104004  and   2013GB111004,
the National Science Foundation of China under grant Nos. 11205132, 11203056  and 11235009,
and Fundamental Research Fund for Chinese Central Universities.

\end{document}